\documentclass[a4paper,11pt,openany]{article}
\usepackage{graphicx}
\usepackage[applemac]{inputenc}
\usepackage{t1enc}
\usepackage{lscape}
\usepackage{amsmath}
\usepackage{amssymb}
\usepackage{cite}
\usepackage{epsfig}
\usepackage{ulem} 
\usepackage [dvipsnames] {xcolor}

\usepackage{xcolor}

\usepackage[colorlinks=true,linkcolor=blue]{hyperref}

\begin{document}

\begin{center}
\Large{\bf Temporal reversibility of reactive systems out of equilibrium:

Molecular dynamics simulation}
\end{center}

\begin{center}
O. Politano$^{a}$, Alejandro L. Garcia$^{b}$, F. Baras$^{a}$,  and M. Malek Mansour$^{c}$
\end{center}

\begin{center}
\small{ (a) Laboratoire Interdisciplinaire Carnot de Bourgogne, $\quad$\\
 UMR 6303 CNRS-Universit\'e Bourgogne Franche-Comt\'e,\\
 F-21078 Dijon Cedex, France

(b) Dept. Physics and Astronomy,\\
 San Jose State University, \\
San Jose, California, 95192 USA

 (c) Universit\'e Libre de Bruxelles CP 231, Campus Plaine,\\
B-1050 Brussels, Belgium}
\end{center}

\date{\today}

\begin{abstract}

The second law of thermodynamics states that entropy production in macroscopic systems is non-negative, reaching zero only at thermodynamic equilibrium. As a corollary, this implies that the state trajectory of macroscopic systems is inherently time-irreversible under out-of-equilibrium conditions. However, over the past half-century, various studies have shown that this principle does not universally apply to the composition sample paths of certain isothermal reactive systems. Theoretical frameworks leading to this surprising observation primarily focus on perfectly homogeneous systems (often referred to as zero-dimensional systems), which inherently exclude the effects of local fluctuations. This oversimplification may account for the paradoxical theoretical predictions. In the absence of relevant experimental data, this paper seeks to explore this phenomenon through microscopic simulations.

\end{abstract}

\section{Introduction}

Consider a perfectly homogeneous, isothermal reactive system in contact with large particle reservoirs.  These reservoirs are assumed to sustain the system in an out-of-equilibrium state by maintaining fixed concentrations of certain reactive components. This type of non-equilibrium systems is commonly generated in laboratory experiments using the so-called "continuously stirred tank reactor" (CSTR) device \cite{Epstein:1998}.  The state of such isothermal systems is solely characterized by its composition, the only pertinent quantity accessible for experimental investigations.

Suppose we are somehow able to measure the precise number of each chemical species over an arbitrarily long time interval. In other words, let us assume that we have access to the exact state trajectory of the system, often referred to as the "path" or "sample path" in the context of stochastic processes. The question arises as to whether the analysis of this state trajectory will allow us to determine the thermodynamic properties of the system. As shown by Evans et al. \cite{Evans:1993},  the answer is affirmative and applies to a large class of non-equilibrium systems. The underlying theory, known as "path thermodynamics", is based on the concept of "path entropy" \cite{Evans:1995,Jarzynski:1997,Lebowitz:1999}. Unlike Boltzmann's entropy, which is defined at a specific instant \cite{Mazur:1984}, path entropy is defined over a time interval of the state trajectory \cite{Crooks:2000,Seifert:2007,Seifert:2005}. Its properties are governed by the so-called "fluctuation theorem" (see \cite{Jarzynski:2011} and \cite{Seifert:2012} for reviews).

While Boltzmann entropy production is strictly non-negative, the fluctuation theorem asserts that the probability of observing a negative path entropy production is not zero, rather it goes as the exponential of the inverse of the number of molecules.  Consequently, this peculiar property has negligible effect on the behavior of macroscopic systems, with its primary relevance lying in extremely small systems. This perspective has fueled significant interest in path thermodynamics descriptions of such systems, leading to notable publications  over the past two decades \cite{Chris:2015} (see  \cite{Chris:2016}for a review).

Despite the extensive literature on the topic, the question of the domain of validity of path thermodynamics has not been fully addressed.  Recently,  we demonstrated that path thermodynamics fails to accurately describe the properties of reactive systems involving more than one elementary reaction leading to the same composition change \cite{Malek:2017,Malek:2020}.  Unsurprisingly, this finding has generated significant skepticism, primarily because our results challenge the universal applicability of the now widely popular path thermodynamics framework to physico-chemical systems.  The main criticisms relate to the use of Markov processes for the mathematical treatment of the problem and the perceived lack of sufficient consideration of system-reservoir interactions.

Recently we identified a simple and realistic set of elementary reactions that conclusively substantiates our claims, without relying on specific assumptions about system-reservoir interactions or requiring complex mathematical developments \cite{Baras:2023b}. This discovery not only resolved superficial controversies but also raised more fundamental questions, to which we will return later. The following straightforward example clearly illustrates the issue.

Let us consider a scenario where a molecule $A$ transforms into a molecule $B$ through interactions (collisions) with either a molecule $S$ or another molecule $B$. Specifically, the reactions are $S + A \rightleftharpoons S + B$ and $B + A \rightleftharpoons 2B$. Starting from an arbitrary state $(A, B \ne 0)$, both reactions lead to the same composition change: $(A, B) \rightarrow (A-1, B+1)$ (forward) or $(A, B) \rightarrow (A+1, B-1)$ (backward). The state trajectory of a reactive system involving these elementary reactions does not contain any information to differentiate between them. Contrary to recent claims \cite{Gaspard:2023}, potential variations in the number of $S$ particles, whether due to other reactive processes or exchanges with reservoirs, do not alter this fact \cite{Baras:2023b}.

From the fundamental principles of irreversible thermodynamics, it is well-established that the entropy production of a reactive system is the sum of the entropy production associated with each individual reaction (see, for example, Section 9.5 of \cite{Kondepudi:2008}). However, this critical information cannot be inferred solely from analyzing the state trajectory. Consequently, the path thermodynamic properties derived from the state trajectory of this system will inevitably be inconsistent with the results of irreversible thermodynamics.

Notably, this observation requires no specific modeling procedures or complex mathematical developments. This conclusively demonstrates the existence of a class of reactive systems that \textit{cannot be described by path thermodynamics}, regardless of how their interactions with reservoirs are managed.

An intriguing consequence of this issue pertains to reactive systems where all elementary reactions lead to the same composition change. Prominent examples include single-variable (denoted $X$) Schlögl-type reactive systems where reactions lead to either $X \rightarrow X+1$ (forward) or $X \rightarrow X-1$ (backward) \cite{Schlgl:1972}. As previously mentioned, the state trajectory lacks information that distinguishes between these reactions. Consequently, its structure is equivalent to that of a system composed of a single reversible reaction, symbolically represented as $X \rightleftharpoons X+1$.  Thus, at the stationary regime, the state trajectory corresponds to that of a thermodynamic equilibrium system. In other words, the state trajectory of Schlögl-type reactive systems is time-reversible in the stationary regime, even when the system is maintained out of equilibrium. However, time-reversal symmetry is a key signature of thermodynamic equilibrium states \cite{Mazur:1984}, which is not necessarily the case here. We are thus faced with a paradox.

It is worth noting that, in contrast to the previous example, the concentration of certain chemicals in Schlögl-type reactive systems is assumed to remain constant. This limitation, related to system-reservoir interactions, could be considered as a potential weakness of the reliability of the result.  We will address this issue later.  Otherwise, no specific modeling procedures or involved mathematical developments are required to establish this rather strange observation.

From a historical perspective, Schnakenberg was the first to highlight the specific characteristics of reactive systems involving multiple elementary reactions that lead to the same change in composition \cite{Schnakenberg:1976}. He developed a network theory to analyze the stationary regime properties of master equations describing reactive systems, with a particular focus on systems maintained out of equilibrium through interactions with reservoirs. Within this framework, a graph is associated with each modification of the system's composition. When multiple elementary reactions lead to the same modification, Schnakenberg's theory requires that the associated graph be decomposed into several subgraphs, each corresponding to one elementary reaction. This theory provides an alternative framework for analyzing the properties of the master equation and proves to be particularly useful in cases involving networks of elementary reactions.

A decade later, using the master equation formulation of reactive systems and Boltzmann's definition of entropy, an analogous analysis was conducted by Luo et al. \cite{Luo:1984}. By implementing a clever rearrangement of transition rates, somewhat similar to Schnakenberg’s procedure, they demonstrated that this framework, referred to as "stochastic thermodynamics" at the time, yielded results that, on average, aligned perfectly with classical thermodynamics. These results were identical to those derived using Schnakenberg's network theory. 

To prevent potential confusion, we will henceforth use the term "path thermodynamics" to denote stochastic thermodynamics based on path entropy, as subtle differences exist between these approaches. For instance, both Schnakenberg's and Luo et al.'s theories pertain exclusively to the master equation itself, whereas the realizations of the corresponding stochastic process are used only in the path thermodynamics approach. As we will demonstrate, this distinction is crucial for understanding the main issue developed in our article.

Two decades later, following the seminal paper by Lebowitz and Spohn \cite{Lebowitz:1999}, Gaspard proposed an exhaustive path thermodynamics description of reactive systems \cite{Gaspard1:2004}. To illustrate his theory, Gaspard used the Schlögl model \cite{Schlgl:1972}, which had been known for over half a century to exhibit temporally reversible stochastic paths, even when the system was maintained out of equilibrium \cite{Graham:1971,VanKampen:1983,Gardiner:2009}. Consequently, this model is fundamentally unsuitable for illustrating specific path thermodynamics properties of a system maintained out of equilibrium.

Rather than merely acknowledging the non-applicability of the theory to such reactive systems, Gaspard employed Schnakenberg’s technique to decompose multi-component paths into their constituent elementary reactions. Together with Andrieux, they successfully demonstrated that this decomposition yields the expected thermodynamic properties of the system \cite{Gaspard2:2004}. However, this approach raises a legitimate question: can it still be considered a true “path thermodynamics” description?

The answer is negative, and the explanation rests on a fundamental property of probability theory: the probability associated with a random event is unique. This principle stems directly from the foundational definition of probability, first introduced by Pascal and subsequently refined by mathematicians, culminating in Kolmogorov's axiomatic framework \cite{VanKampen:1983,Gardiner:2009}.

Consider, for instance, a reactive system described by a jump Markov process $\chi(t)$. The probability of observing a transition during a brief time interval $\Delta t$, from the state $\chi(t) = X'$ to the state $\chi(t + \Delta t) = X$, is given by the conditional probability $P(X, t + \Delta t \, | \, X', t)$, representing the likelihood of the system being in state $X$ at time $t + \Delta t$, given it was in state $X'$ at time $t$. However, a transition $X' \rightarrow X$ is fundamentally a random event, and therefore, its probability must be unique. In addition, since $\chi(t)$ is governed by a jump Markov process, the Kolmogorov equality ensures that, to dominant order in $\Delta t$, $P(X, t + \Delta t \, | \, X', t) = W(X \, | \, X') \Delta t$, where $W$ denotes the transition rate (conditional probability per unit time) \cite{VanKampen:1983,Gardiner:2009}. Consequently, the transition rate $W$ must also be unique.

In particular, if a reactive system involves $\rho$ elementary reactions leading to the same change in composition, the resulting transition rate must necessarily be the sum of the transition rates associated with each reaction. Specifically, $W(X \, | \, X') = \sum_{\rho} W_\rho(X \, | \, X')$, where $W_\rho(X \, | \, X')$ represents the transition rate associated with the individual elementary reaction $\rho$. As a result, expressing the sample path of $\chi(t)$ in terms of {\it individual} transition rates $W_\rho(X \, | \, X)$ leads to a peculiar kind of state trajectory that contradicts the fundamental properties of any known stochastic process at the level of their sample paths. 

With other arguments falling short, a different line of criticism was recently directed at our findings, suggesting that multiple jump Markov processes could potentially model a given reactive system, thereby challenging the uniqueness of stochastic formulations for non-equilibrium systems \cite{Gaspard:2021,Gaspard:2023}. Beyond the mathematical proofs that refute this claim \cite{Baras:2023,Baras:2023b}, as detailed in the last two paragraphs, we also conducted microscopic simulations of reactive Boltzmann equations to provide further comprehensive evidence, employing the well-established Bird algorithm \cite{Bird:1976,Garcia:2022}.  This approach, which is significantly faster than traditional molecular dynamics simulations, enables us to simulate perfectly homogeneous dilute gas systems while preserving key aspects of microscopic dynamics. As demonstrated in \cite{Baras:2023}, these simulations consistently validate the predictions of the standard stochastic modeling of reactive systems \cite{Malek:2017,Malek:2020}.

Driving a system out of equilibrium necessarily requires the presence of reservoirs and external forces that can be switched on and off to closely replicate laboratory conditions. In the standard stochastic modeling, established over half a century ago \cite{VanKampen:1983, Gardiner:2009}, the state of the reservoirs and the properties of external forces are predefined. Their role is thus limited to maintaining constant given system parameters, such as temperature or the concentration of certain chemical components.

This article, along with all of our related previous works, is based on the standard stochastic modeling of reactive systems.  It is worth noting that nearly everything we know about the statistical properties of non-equilibrium systems has been obtained from this type of modeling, like for exemple the Landau-Lifshitz fluctuating hydrodynamics formalism \cite{Land-Stat} or the onset of instabilities in nonequilibrium systems \cite{Nicolis-Haken}.   But the most striking example is undoubtedly the use of standard stochastic modeling  by Lebowitz and Spohn to set up the basics of path thermodynamic formulation of reactive systems \cite{Lebowitz:1999}.  The same type of modeling that restricts the applicability of path thermodynamics to reactive systems where each observed change in composition corresponds to one, and only one, elementary reaction.  As shown in \cite{Malek:2020,Baras:2023}, the  paradoxical behavior of Schl\"{o}gl-type reactive systems is a direct consequence of this approach.

The question, therefore, arises: are these unexpected properties of non-equilibrium state trajectories a genuine feature of reality, or are they merely a consequence of an oversimplified treatment of system-reservoir interactions in standard stochastic modeling?

As Schmied and Seifert have noted \cite{Seifert:2007}, an accurate description of reactive systems should, in principle, incorporate explicit interactions with reservoirs, resulting in a unified system-reservoir entity. However, the theoretical approach underlying this course of procedure is quite complex and, to date, have been applied only in relatively simple scenarios. One such example is the explicit inclusion of the degrees of freedom of heat reservoirs within the analysis using a Hamiltonian framework at the microscopic level. This approach was primarily advanced by Jarzynski, who successfully derived fundamental results related to various thermodynamic aspects of the fluctuation theorem \cite{Jarzynski:1997}.

The case of single-variable Schl\"{o}gl-type reactive systems presents a particular challenge, as at least one of the elementary reactions must involve a chemical species shared with the reservoirs in order to maitain non-equilibrium constraints. Detailed studies indicate that explicitly including reservoir interactions to form a unified system-reservoir entity is fraught with difficulties and is far from straightforward. As discussed in \cite{Baras:2023,Baras:2023b}, a primary conceptual hurdle is that the system-reservoir entity itself becomes a closed system. Unless infinitely large reservoirs are assumed, this compound system will inevitably reach a thermodynamic equilibrium state over time. At present, there is no clear method to maintain such a system-reservoir entity in a non-equilibrium state over extended periods.

On the other hand, it is well understood that the time-reversibility of Schlögl-type reactive systems is a direct consequence of path thermodynamics' inability to accurately describe the properties of reactive systems involving more than one elementary reaction leading to the same composition change \cite{Malek:2017,Malek:2020}. As noted above, the validity of this fact is unaffected by how system-reservoir interactions are handled. Although there are quite plausible presumptions supporting the validity of standard Markovian modeling for reactive systems, the current state of our knowledge does not yet provide a definitive answer to this question.

As was shown above by our first example, the validity of this fact is unaffected by how system-reservoir interactions are handled. Although there are quite plausible presumptions supporting the validity of standard Markovian modeling for reactive systems, the current state of our knowledge does not yet provide a definitive answer to this question.

There remains one last possibility that should be considered to either rule out or at least explain this paradoxical result. This pertains to the fact that the general framework leading to this peculiar result concerns "ideally homogeneous" systems (often referred to as zero-dimensional systems) that exclude the presence of local fluctuations. However, there exist numerous situations where local fluctuations have dramatic effects on the macroscopic behavior of reactive systems. Known examples include annihilation reactions in low-dimensional systems \cite{Lindenberg:1995,Baras:2004} or the onset of instabilities in chemical systems where local fluctuations destroy coherent oscillatory behavior (limit cycle) in one and two-dimensional systems \cite{Baras:1996, Malek:2000}. The same mechanism may perhaps provide an explanation for the perceived paradoxical issue in Schl{\"o}gl-type reactive systems.  In the absence of pertinent experimental results, another way to explore this possibility is to resort to classical molecular dynamics simulation of reactive systems. 

\section{Hard sphere simulations of reactive systems}

Generating statistically reliable data to estimate two-body distribution functions poses a  challenge due to the need for a large number of data points. One approach to mitigate this challenge is by conducting molecular dynamics simulations of dilute hard sphere gases, where careful treatment of the collision timetable can greatly enhance computational efficiency \cite{Rapaport:2004}. However, it's important to note that highly efficient algorithms may introduce unintended biases, such as Maxwell's demon, in certain scenarios.  In this study, we followed the approach outlined in \cite{Rapaport:1980} to address these challenges. To eliminate potential sources of errors, we rebuilt the collision timetable completely every 200 steps. This procedure was tested through a simulation involving $10^6$ collisions, and the results were found to be comparable to those obtained using the standard procedure.

Another advantage of choosing a system of dilute hard spheres is that the results can be compared  with those generated through microscopic simulations of the homogeneous reactive Boltzmann equation. The optimal strategy would be to conduct a molecular dynamic simulations of the same reactive systems that were considered previously \cite{Baras:2023}. This is precisely the approach adopted in the present work. Accordingly, we consider the following models:
\begin{equation}
\label{Eq1}
A \,\, + \,\, X \,\,\,  \mathop{\rightleftharpoons}^{k_1}_{k_{- 1}} \, \,\, X \,\, + \,\, X  \, \,\, ; \, \,\, B \,\, + \,\, C \,\,\,  \mathop{\rightleftharpoons}^{k_2}_{k_{- 2}} \, \,\, B \, + \, X
\end{equation}
and
\begin{equation}
\label{Eq2}
A \,\, + \,\, A \,\,\,  \mathop{\rightleftharpoons}^{k_1}_{k_{- 1}} \, \,\, X \,\, + \,\, X  \, \,\, ; \, \,\, B \,\, + \,\, C \,\,\,  \mathop{\rightleftharpoons}^{k_2}_{k_{- 2}} \, \,\, B \, + \, X
\end{equation}
where the mole fraction of reactants $A$, $B$, and $C$ is assumed to remain constant.  Their values, as well as those of the kinetic constant $k_{\pm \, i}$, $i = 1, 2$, are set in order to ensure that both systems operate under non-equilibrium conditions \cite{Baras:2023}.

More precisely, for the model (\ref{Eq1}) we set $k_{-1} = k_{1}$, $k_{-2} = k_{2}$, $C = A / 2$, and $A / B = k_{2} / k_{1} = 5 / 6$, so that the macroscopic number of $X$ particles at the stationary state reads $\overline{X}_s = A$.  This choice of parameters guaranties that the system operates under non-equilibrium conditions.  For instance, for these parameters we can check that in dilute (ideal) systems, the {\it thermodynamic entropy production}, $\sigma_s$, is strictly positive at the stationary state
\begin{equation}
\label{Eq3}
\sigma_s \, = \, \frac{6}{5} \, N \, k_B \, k_2 \, a^2 \,  \ln(2) \,\, > \,\, 0
\end{equation}
where $k_B$ is the Boltzmann constant, $a$ the mole fraction of $A$, and $N$ the total number of particles present in the system, including solvent or other non-reactive particles (extensivity parameter) \cite{Kondepudi:2008}.

For model (\ref{Eq2}), we shall use a slightly different parameter setting: $k_1 = k_2 = k_{-1} = k_{- 2}$, $B = 5 \, A /3$ and $C = A /3$. The main motivation for this parameter choice is to ensure that the macroscopic number of $X$ particles at the stationary state matches that of system (\ref{Eq1}), namely $\overline{X}_s = A$. This will make the comparison of simulation results between these two models more straightforward.  Furthermore,  here again we can check that in dilute (ideal) systems the thermodynamic entropy production is strictly positive at the stationary state:
\begin{equation}
\label{Eq4}
\sigma_s \, = \, \frac{5}{9} \, N \, k_B \, k_1 \, a^2 \,  \ln(9/2) \,\, > \,\, 0
\end{equation}
so that the system operates under non-equilibrium conditions.

We observe that in the reactive system described by (\ref{Eq2}), each elementary reaction results in a distinct change in composition, rendering the associated state trajectory inherently time-irreversible \cite{Kondepudi:2008}. In contrast, the reactive system (\ref{Eq1}) is characterized by both reactions leading to either $X \rightarrow X+1$ (forward) or $X \rightarrow X-1$ (backward). Consequently, in an ideally homogeneous system, the state trajectory of this reactive system remains time-reversible in non-equilibrium stationary regimes \cite{Malek:2017,Malek:2020,Baras:2023}. The primary objective of this article is to determine whether this paradoxical property is an intrinsic feature of the system or a consequence of the ideal homogeneity (zero-dimensional) assumption. Comparing the statistical properties of these two systems will play a decisive role in interpreting the results of the molecular dynamics simulations.

Another issue is how are we going to proceed in order to maintain constant the mole fraction of a chemical species.  This problem represented a serious challenge in early-day microscopic simulations of reactive systems.  Achieving this required bringing the system into contact with a reservoir containing that species and adjusting the parameters so that the frequency of exchange processes with the reservoir was much higher than that of the reactive processes. However, this approach resulted in significant CPU wastage, making detailed studies of reactive processes through microscopic simulations prohibitively expensive.

In the mid-1970s, an original solution to tackle this problem was proposed \cite{Portnow:1975,Boissonade:1979}. Here's how it works: In order to maintain the mole fraction of $A$ and $C$ constant ($B$ is not modified), "solvent" particles, designated as $S$, are introduced, and the following strategy is implemented: when a reactive collision results in the destruction of one of these particles, an $S$ particle is randomly selected from the immediate vicinity of the collision center and transformed into the destroyed particle. Likewise, the same strategy for the reverse process is applied. When an $A$ or $C$ particle is generated through a reactive collision, the aforementioned strategy is used to select one such particle and replace it with an $S$ particle (see \cite{Baras:1997} for a review). This is the optimal strategy for traditional molecular dynamic simulation of reactive systems ($\rm{MD}$).

An alternative option is to randomly select an $S$ particle from the entire system volume and transform it back into an $A$ or $C$ species each time they are destroyed through a reactive process. Similarly, if one of these particles is generated through a reactive collision, that same type of particle is randomly selected from the entire system volume and transformed into an $S$ particle. This approach aims to reduce the local fluctuations of $A$ and $C$ particles by promoting their spatial homogenization. It can be seen as a hybrid method, combining elements of traditional molecular dynamics and purely homogeneous Bird simulation procedures. We refer to this alternative option as "hybrid simulation" ($\rm{HS}$). The distinction between these two choices, as well as the strictly homogeneous Bird procedure, should offer valuable insights into the impact of local fluctuations on the system's statistical properties.

For microscopic modeling of reactive processes we adopt the "hard sphere chemistry" strategy, which originated in the mid-1970s \cite{Portnow:1975,Boissonade:1979}. The fundamental concept is elegant and straightforward. First, we assign each particle a distinct attribute, such as a "color". If two particles collide with sufficient energy, meaning that their relative kinetic energy exceeds a threshold linked to the activation energy of the reaction, a reactive collision takes place \cite{Present:1959}. In this event, the particles' attributes (colors) are changed according to the chemical step in question.

A significant issue with this approach is the inevitable deformation of the Maxwell-Boltzmann distribution, as only the most energetic particles can undergo a reactive transformation \cite{Prigogine:1949,Baras:1989}.  In order to prevent this non-equilibrium effect, the reactive collision rules have to be further simplified.  Consider an isothermal bi-molecular chemical reaction, say $R_i$, similar to one of the forward or backward reactions considered above,    and let us denote by $E_i$ the associated activation energy.  Upon expressing the composition of the system in terms of mole fractions, it can easily be shown that the kinetic constant $k_i$ of that reaction $R_i$ is proportional to the collision frequency $\nu_i$ of the reactant particles, that is $k_i  =  \nu_i \, \widetilde{k}_i$, where $\widetilde{k}_i = \exp \big\{- \, E_i / k_B \, T \big\}$ stands for the Arrhenius factor ($T$ represents the temperature of the medium and $k_B$ is the Boltzmann constant).  After a collision between two such reactive particles has occurred, we choose randomly $\widetilde{k}_i \%$ of the collisions to be reactive.   This well known procedure avoids the deformation of the Maxwell-Boltzmann distribution since it does not involve any systematic energy transfer between reactants and products.  It is, however, restricted to isothermal second-order (binary collisions) reactions (see \cite{Baras:1997} for a review).

\section{Simulation results}

For microscopic simulations, we consider an ensemble of $N$ hard spheres with identical mass $m$ and diameter $d$,  regardless of their chemical identity. They are confined within a volume $V$ chosen to yield a number density of $3 \times 10^{-3}$ particles per cubic diameter $d^3$. This choice ensures that the system falls within the valid range of the Boltzmann equation. Further details can be found in references \cite{Bird:1976, Garcia:2022}.

For the microscopic simulation of the reactive system (\ref{Eq1}), the total number of hard spheres is set to  $N = 5,000$, with $A = 1,000$, $B = 1,200$, and $C = 500$ particles. The rate constants are set to $k_1 = k_{-1} = 0.9 \nu$ and $k_2 = k_{-2} = 0.75 \nu$, where $\nu$ is the collision frequency.  In other words, $90 \%$ of  collisions between particles involved in the first reaction and $75 \%$ of the collisions between particles involved in the second reaction are assumed to be reactive. We can readily confirm that this parameter set results in a macroscopic number of $X$ particles equal to $\overline{X}_s = 1,000$ in the stationary state.

For the microscopic simulation of the reactive system (\ref{Eq2}),  we set the value of $N$ to be $N = 7,000$.  The other parameters are set as follows: $A = 1,500$, $B = 2,500$, and $C = 500$ particles, with $k_1 = k_2 = k_{-1} = k_{-2} = 0.9 \nu$.  In other words, $90 \%$ of  collisions between particles involved in the first and in the second reactions of the model  (\ref{Eq2}) are assumed to be reactive. With these parameter choices, the macroscopic number of $X$ particles in the stationary state aligns with that of the reactive system (\ref{Eq1}), i.e.,  $\overline{X}_s = 1,000$.

Our primary objective in conducting these simulations is to analyze the statistical properties of sample paths associated with the reactive systems (\ref{Eq1}) and (\ref{Eq2}). By measuring the number of $X$ particles over a sufficiently long time interval, we can estimate the "direct" probability distribution $P(X_1, t_1 \, ; \, X_2, t_2)$ of having $X_1$ particles at time $t_1$ and $X_2$ particles at time $t_2 \, > \, t_1$, which can then be compared with the "reverse" probability distribution $P(X_2, t_1 \, ; \, X_1, t_2)$ of having $X_2$ particles at $t_1$ and $X_1$ particles at $t_2$. Special care is taken to ensure that, before starting the collection of data, the system has reached the stationary regime. The resulting (estimated) probability distributions therefore depend only on the time interval $\tau = t_2 - t_1$.

We observed that the statistical errors associated with these probability distributions grow quite rapidly for time intervals $\tau$ larger than $20$ mean reactive collision times (MRCT). Therefore, we limit the simulation to values of $\tau$ not exceeding $15$ MRCT. Furthermore, for practical reasons, we set the value of $X$ at the final time $t_2 = t_1 + \tau$ to a "reference state" $X_{\rm ref}$. As can be expected, we observed that the statistical error decreases as the chosen value of $X_{\rm ref}$ approaches the most probable state. Given that the most probable state is located in close proximity to the macroscopic stationary state $\overline{X}_s = 1000$, we set $X_{\rm ref} = \overline{X}_s$.

To carry out our measurements, we generated four different datasets, each containing outcomes associated with $10^{10}$ collisions.   The resulting direct and reverse probability distributions are shown in Figure (1) for both molecular dynamics ($\rm{MD}$) and hybrid simulation ($\rm{HS}$) cases, where $\tau = 10$ mean reactive collision time. The statistical error, estimated from 20 successive runs of $5 \times 10^8$ collisions, does not exceed $1\%$ (about by the size of the marker symbols in Figure (1)).

As expected, the sample paths associated with model (\ref{Eq2}) are distinctly time-irreversible. Furthermore, no notable differences between molecular dynamics and hybrid simulations are observed. A property that holds true for perfectly homogeneous (Bird) simulation results as well. However, this proves to be not the case for model (\ref{Eq1}), where quantitative discrepancies between ($\rm{MD}$)  and ($\rm{HS}$) results are apparent.

Quite surprisingly, these discrepancies observed in model (1) are only quantitative, meaning that the qualitative behavior remains the same in both cases.  As shown in Figure (1), the direct and reverse trajectories are practically indistinguishable in both the case of (MD) and the case of (HS).  In other words, $P(X, t \, ; \, X_{\text{ref}}, t + \tau) \approx P(X_{\text{ref}}, t \, ; \, X, t + \tau)$, well within the limits of statistical errors, despite the fact that the system is \emph{not} at equilibrium.

One way to highlight this property is to consider the ratio of the direct and reverse probabilities.  As shown in Figure (1), this ratio is practically equal to one for the model (1), but not for the model (2).   This observation is further supported in Figure (2), where the ratio $P(X, t \, ; \, X_{\text{ref}}, t + \tau) / P(X_{\text{ref}}, t \, ; \, X, t + \tau)$ is shown.

\section{Concluding remarks}

The microscopic simulation results unequivocally demonstrate that the sample paths of a reactive system can exhibit time-reversibility, even when the system is deliberately maintained out of equilibrium (model (1)). We acknowledge our perplexity and the lack of entirely satisfactory explanations for this observed phenomenon.  Given the unexpected nature of these results, it is useful to briefly outline the fundamental principles of microscopic molecular dynamics simulations for physico-chemical systems, as well as the specific methodologies we have applied to reactive systems. This may help to rule out any inaccuracies that could have contributed to the unforeseen findings.

Microscopic molecular dynamics simulations of physico-chemical systems typically involve the system interacting with large external reservoirs that help maintaining certain quantities, such as temperature or pressure, at constant values. To achieve this, the simulation parameters are set so that the exchange processes with the reservoirs occur at a much higher frequency than the system's internal processes. This setup, however, often results in considerable computational inefficiencies. Historically, several approaches have been developed to address these inefficiencies, some of which parallel methods used in standard stochastic modeling \cite{Frenkel:2001}. A well-known example is the Nose-Hoover "thermostating" scheme, which employs sophisticated numerical techniques to avoid the direct simulation of the extensive number of degrees of freedom associated with the reservoir itself \cite{NosHov:1984}.

Comparable numerical techniques have also been created specifically for the microscopic simulation of reactive systems.  The primary aim of these efforts has been to develop an efficient method for regulating the concentration of specific products, while avoiding an excessive increase in the computational demands of the numerical procedure.  Obviously, skepticism might lead one to question the validity of the procedure that we have used to maintain constant concentrations of specific products, achieved by introducing solvent particles $S$.

This methodology, pioneered by Boissonade in $1979$ \cite{Boissonade:1979}, has since been widely adopted in the simulation of various reactive systems. These include systems exhibiting sustained chemical oscillations or bimodal instabilities \cite{Baras:1990,Baras:1996,Malek:2000}, highly exothermic (explosive) reactions \cite{Baras:1989}, wave propagation phenomena \cite{Lemarchand:1999,Lemarchand:2006}, and annihilation reactions in low-dimensional systems \cite{Baras:2004}. To date, no challenges have been raised regarding the legitimacy of this technique, nor have alternative methods been proposed.

Finally, one might question whether this unexpected result is limited to specific categories of reactive systems or if other non-equilibrium systems could exhibit similar behavior. In this respect, two cases have recently caught our attention: fluid mixtures under concentration gradient and solids  under temperature gradient.  

Ongoing research in this area suggests that unexpected scenarios may arise, particularly in the
case of a fluid mixture under a concentration gradient. Theoretical findings based on both the
fluctuating hydrodynamics formalism and random walk modelling of particle diffusion (master
equation formulation) yield results that are strikingly similar to those presented in this article.
We are currently conducting extensive molecular dynamics simulations to validate the accuracy
and relevance of these theoretical predictions.

\begin{figure}[h!]
\begin{center}
\epsfclipon
\epsfig{file=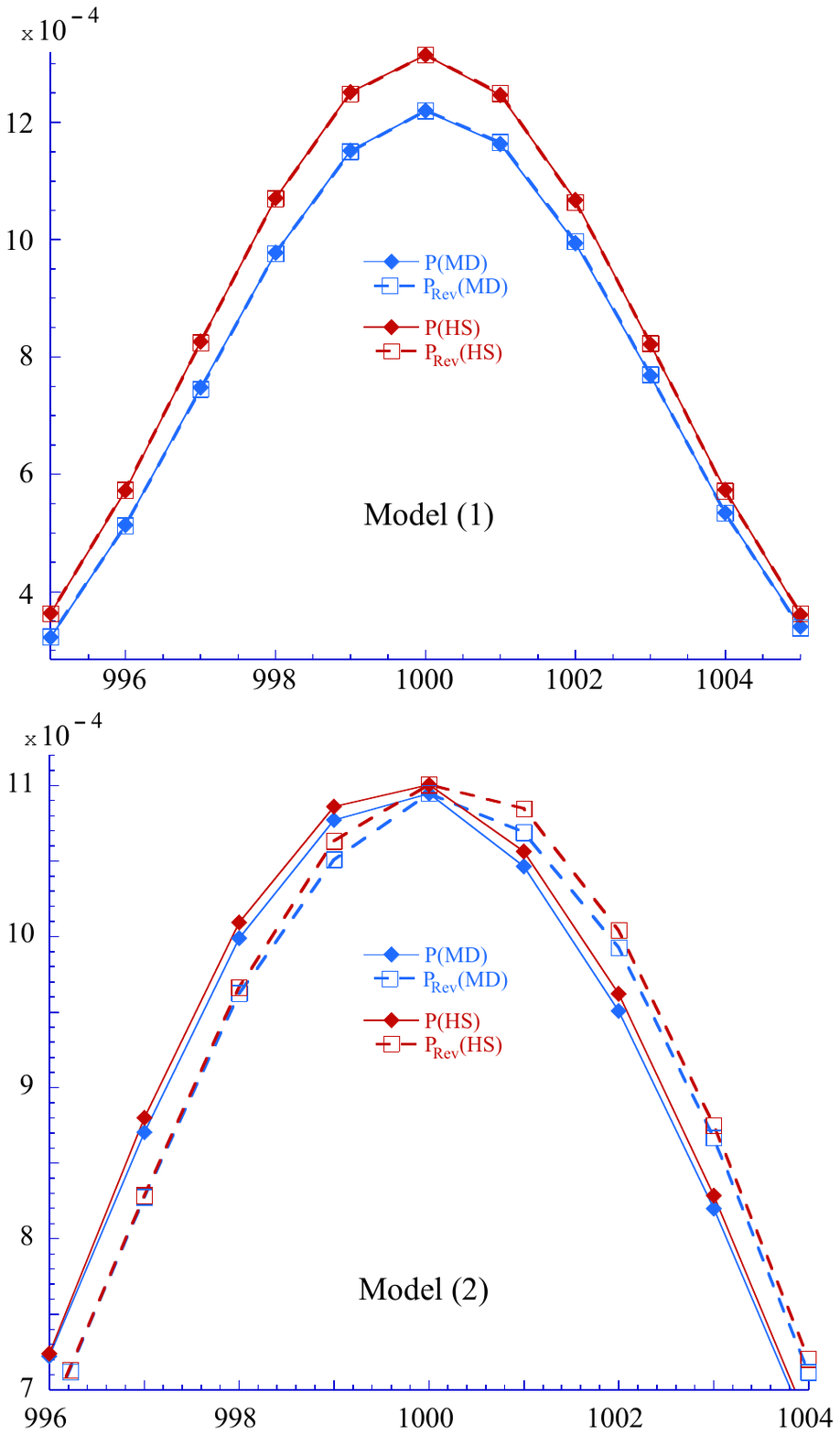,width=14.cm}
\caption{\label{fig:Probs}\small{Direct probability distribution $P(X, t \,\, ; \, X_{ref}, t + \tau)$ (diamonds-solid lines)
and reverse probability distribution $P(X_{ref}, t \,\, ; \, X, t + \tau)$ (squares-dashed lines)
versus $X$ for: (top) model (\ref{Eq1});  (bottom) model (\ref{Eq2}).
The macroscopic stationary state is $X_{ref} = 1000$ and  $\tau = 10$ mean reactive collision time, as measured in the simulations.
Results obtained by (MD) are in blue and those from (HS) are in red.
}}

\label{figOne}
\end{center}
\end{figure}

\begin{figure}[h!]
\begin{center}
\epsfclipon
\epsfig{file=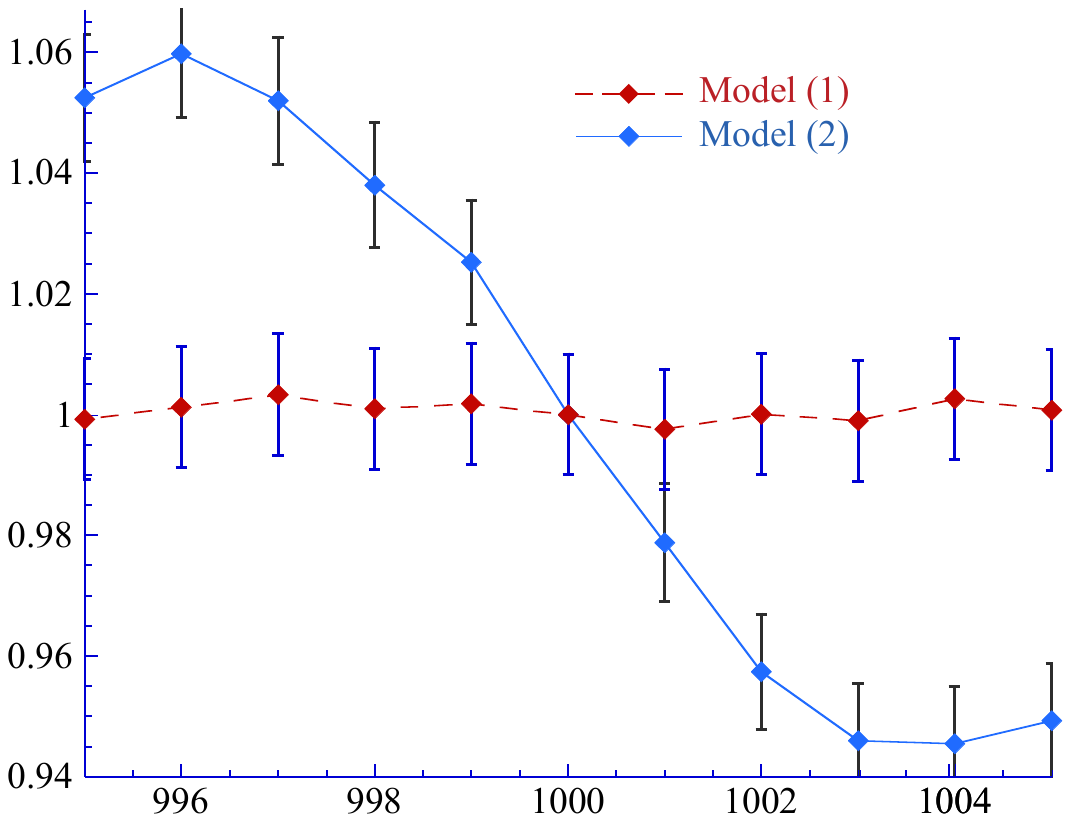,width=14.cm}
\caption{\label{fig:ProbRatio}\small{Probability ratio $P(X, t \,\, ; \, X_{ref}, t + \tau) \, / \, P(X_{ref}, t \,\, ; \, X, t + \tau)$ obtained throug ({\rm MD}) procedure {(see the caption of Fig.~1).}  }}
\label{figTwo}
\end{center}
\end{figure}

\section*{Acknowledgments}
The use of computational facilities at the Computing Center of the University of Bourgogne, DNUM-CCUB, is gratefully acknowledged.  One author (AG) acknowledges support by the U.S. Department of Energy, Office of Science, Office of Advanced Scientific Computing Research, Applied Mathematics Program under contract No. DE-AC02-05CH11231.

\end{document}